\documentclass[12pt]{article}
\usepackage{amsmath}
\usepackage{epsf}
\usepackage{epsfig}
\usepackage{here}
\usepackage{amssymb}
\usepackage{citesort}
\usepackage{graphicx}
\usepackage{latexsym}
\newcommand{\be}{\begin{equation}}
\newcommand{\ee}{\end{equation}}
\newcommand{\bear}{\begin{eqnarray}}
\newcommand{\ear}{\end{eqnarray}}

\newsavebox{\LSIM}
\sbox{\LSIM}{\raisebox{-1ex}{$\ \stackrel{\textstyle<}{\sim}\ $}}

\newsavebox{\GSIM}
\sbox{\GSIM}{\raisebox{-1ex}{$\ \stackrel{\textstyle>}{\sim}\ $}}

\begin{document}
\begin{titlepage}
\begin{flushright}
hep-ph/0211056
\end{flushright}
$\mbox{ }$
\vspace{.1cm}
\begin{center}
\vspace{.5cm}
{\bf\Large Flavor Physics and Warped Extra Dimensions\footnote{Talk presented at SUSY02, DESY Hamburg, Germany, June 17-23, 2002}}\\
\vspace{1cm}
Stephan J. Huber\footnote{stephan.huber@desy.de} \\ 
 
\vspace{1cm} {\em Deutsches Elektronen-Synchrotron DESY, Hamburg, Germany}\\[.2cm]

\end{center}
\bigskip\noindent
\vspace{1.cm}
\begin{abstract}
We consider a five dimensional model with warped geometry 
where the standard model fermions and gauge bosons correspond 
to bulk fields. Fermion masses and CKM mixings can be explained 
in a geometrical picture, without hierarchical Yukawa couplings. 
We discuss various flavor violating processes induced by 
(excited) gauge boson exchange and non-renormalizable operators.
Some of them, such as muon-electron conversion, are in the
reach of next generation experiments.
\end{abstract}
\end{titlepage}
\section{Introduction}
The huge discrepancy between the Planck scale, 
$M_{\rm Pl}\sim10^{19}$ GeV, and the scale of
electroweak symmetry breaking, $M_W\sim10^{2}$ GeV,
is one of the most interesting challenges in modern
physics.  Recently, it was realized
that a small but warped extra dimension provides
an elegant solution to this gauge hierarchy 
problem \cite{RS}. The fifth dimension is an $S_1/Z_2$
orbifold with an AdS$_5$ geometry, bordered
by two 3-branes with opposite tensions and separated
by distance $R$.  The 
AdS warp factor $\Omega=e^{-\pi k R}$
generates an exponential hierarchy between the
effective mass scales on the two branes.
If the brane separation is $kR\simeq 11$, the scale on 
the negative tension brane is of TeV-size, while the scale on 
the other brane is of order $M_{\rm Pl}$.  The 
AdS curvature $k$ and the 5D Planck mass $M_5$ 
are both assumed to be of order $M_{\rm Pl}$. At the 
TeV-brane gravity is weak because the zero mode
corresponding to the 4D graviton is localized at the
positive tension brane (Planck-brane). 

We take the SM gauge bosons and fermions as bulk fields. 
The Higgs field is localized at the TeV-brane,
otherwise the gauge hierarchy problem would reappear \cite{CHNOY,HS}. 
Comparison with electroweak data, in particular with the weak 
mixing angle and gauge boson masses, requires the KK 
excitations of SM particles to be heavier than about 10 TeV \cite{HS,HLS,HPR}. 
If the fermions were confined to the TeV-brane, the KK scale 
would be even more constrained.

Models with localized gravity open up attractive possibilities for flavor 
physics. If the SM fermions reside in the 5-dimensional bulk, the 
hierarchy of quark and lepton masses can be interpreted in a
geometrical way \cite{GP,HS2}. Different flavors are localized at 
different positions in the extra dimension or, more precisely, 
have different wave functions. The fermion masses are in direct 
proportion to the overlap of their wave functions with the Higgs field \cite{AS}.
Also the CKM mixing can be explained along these lines. 
Moreover, bulk fermions reduce the impact of non-renormalizable 
operators which, for instance, induce flavor violation and rapid 
proton decay, since closer to the Planck-brane the effective 
cut-off scale of the model is enhanced \cite{GP,HS2}. 
Small Majorana neutrino masses
can then arise from dimension five interactions without
introducing new degrees of freedom \cite{HS4}. The atmospheric 
and solar neutrino anomalies can be satisfactorily resolved.
Alternatively, Dirac neutrino masses can generated by a coupling to
right-handed neutrinos in the bulk \cite{GP,HS3}.
 
In this talk we review how fermion masses and mixings 
can be related to a ``geography'' of fermion locations in
the extra dimension. Flavor violation by (excited) gauge boson 
exchange is a natural consequence of this approach. 
Contributions from non-renormalizable operators turn out to
be safely suppressed. We discuss various flavor violating 
processes, especially focusing on the lepton sector. Some of 
them, such as muon-electron conversion, are in the reach of 
next generation experiments.

\section{Fermion masses and CKM mixings}
By the Kaluza-Klein (KK) procedure the 5D fields are decomposed
into an infinite tower of 4D fields. The wave functions encode 
information on where the KK states are localized in the extra dimension.
Together with the KK masses they are obtained by solving the
5D equations of motion. In five dimensions fermions are vector-like,
and we can associate with them a 5D Dirac mass term, parameterized 
by $m_{\Psi}=c\cdot k ~{\rm sgn}(y)$, where $y$ is the 5th coordinate. 
Depending on the $Z_2$ orbifold transformation property of the fermion, 
$\Psi(-y)_{\pm}=\pm \gamma_5 \Psi(y)_{\pm}$,
the left-handed (right-handed) zero mode, $f_0\sim e^{(2-c)k|y|}$, 
of the KK decomposition is projected out by the boundary conditions \cite{GN,GP,HS2}.
The KK excited states come in left- and right-handed pairs, which 
are degenerate in mass. Note that the 5D Dirac mass parameter regulates 
whether the zero mode is localized towards the Planck-brane $(c>1/2)$
or the TeV-brane $(c<1/2)$.
   
\begin{figure}[t] 
\begin{picture}(100,145)
\put(85,-10){\epsfxsize7cm \epsffile{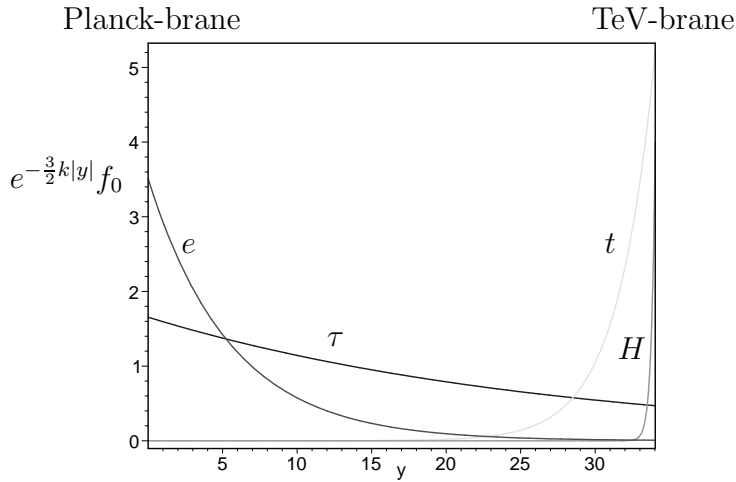}}
\put(40,100){{$e^{-\frac{3}{2}k|y|}f_0$}} 
\put(105,75){{$e$}} 
\put(160,40){{$\tau$}} 
\put(265,75){{$t$}} 
\put(270,35){{$H$}} 
\put(260,160){{TeV-brane}} 
\put(60,160){{Planck-brane}} 
\end{picture} 
\caption{Localization of the electron, tau 
and top quark zero modes in the fifth dimension
together with the Higgs profile ($y$ is given in units of $k$).}
\label{f_1}
\end{figure}

Masses for the fermionic zero modes, which are associated
with the SM quarks and leptons, are generated by the Higgs mechanism.
The induced fermion masses 
\begin{equation}
M_{ij}=\int_{-\pi R}^{\pi R}\frac{dy}{2\pi R}\lambda^{(5)}_{ij}
             e^{-4\sigma}H(y) f_{0L}^{(i)}(y)f_{0R}^{(j)}(y)
\end{equation}
crucially depend on the overlap between the Higgs and fermion wave
functions in the extra dimension, and naturally become small
for fermions residing close to the Planck-brane.
Here $\lambda^{(5)}_{ij}$ are the 5D Yukawa couplings, $H$ is
the Higgs profile localized at the TeV-brane, and $f_0^{(j)}$ are 
the zero modes of the relevant quark and lepton fields. In fig.~\ref{f_1}
we sketch this five dimensional ``geography''. 
Assuming non-hierarchical 5D Yukawa couplings
of the order of the 5D weak gauge coupling $g_2^{(5)}$, the localized
Higgs field induces a product-like pattern for the mass matrices
\begin{equation} \label{product}
M\sim\left(\begin{array}{ccc} 
a_1b_1 &a_1b_2  & a_1b_3 \\
a_2b_1 &a_2b_2  & a_2b_3 \\
a_3b_1 &a_3b_2  & a_3b_3 \\
\end{array}\right)
\end{equation}
where $a_i$ and $b_i$ depend exponentially on the associated $c$
parameters. If the mass matrix is diagonalized by $U_LMU_R^{\dagger}$,
the left- and right-handed mixings are typically $U_{L,ij}\sim a_i/a_j$
and $U_{R,ij}\sim b_i/b_j$, respectively. Only fermions which have
similar positions ($c$ parameters) have large mixings. The mass matrix
(\ref{product}) predicts the approximate relation $U_{13}\sim U_{12}U_{23}$
between the mixing angles, which in case of the observed CKM matrix
is satisfied up to a factor of about two \cite{PDG}.

Building up the fermion mass matrices from eq.~(\ref{product})
requires the specification of $c$ parameters and Yukawa couplings.
Thus there are considerably more independent parameters in the model
than there are observable fermion masses and mixings. In the following
we assume that the pattern of fermion masses is, at the first place, 
determined by the fermion locations. We are looking for a set of
$c$ parameters, where typical Yukawa couplings $\lambda^{(5)}_{ij}\sim g_2^{(5)}$
reproduce the observed fermion properties. More precisely, we are
generating random sets of Yukawa couplings and require the averaged
fermion masses and mixings to fit the experimental data. Taking
$1/\sqrt{2}<|\lambda^{(5)}_{ij}/g_2^{(5)}|<\sqrt{2}$ and random phases
from 0 to $2 \pi$, we find the ``optimal'' locations
\begin{eqnarray}
c_{Q1}=0.65, \quad &c_{D1}=0.65, \quad  &c_{U1}=0.67, \quad  \nonumber \\
c_{Q2}=0.59, \quad &c_{D2}=0.61, \quad &c_{U2}=0.53, \quad\nonumber \\
c_{Q3}=0.32, \quad &c_{D3}=0.61, \quad &c_{U3}=-0.60. \quad  \label{ps}
\end{eqnarray}
With exception of $V_{ub}$ which is too large by a factor of two, all
quark masses and mixings are on average within their allowed ranges.
As expected from their similar locations, the right-handed rotations of 
the down quarks turn out to be large. Note that (\ref{ps}) is different
from the locations we used in ref.~\cite{HS2} to maximally
suppress proton decay. There is a degeneracy in the solution (\ref{ps})  
since the fermion masses do not change if the all left- and right-handed 
quarks are shifted oppositely by the same amount. The quarks can be 
localized closer towards the Planck-brane by $\delta c=\ln(l)/(2\pi R)$ if the 5D 
Yukawa couplings are increased by a common factor $l$. 

To determine the lepton locations we have to take into account neutrino
masses and mixings. We assume that dimension five interactions induce
small Majorana neutrino masses \cite{HS4}. Large neutrino mixings require the
neutrinos and thus the SU(2) lepton doublets to have similar positions $c_{Li}$.
To suppress the matrix element $U_{e3}$  it is favorable to separate the
electron doublet somewhat from the muon and tau doublets 
 \begin{eqnarray}
c_{L1}=0.63, \quad &c_{L2}=0.58, \quad  &c_{L3}=0.58, \quad  \nonumber \\
c_{E1}=0.75, \quad &c_{E2}=0.62, \quad &c_{E3}=0.50. \label{psl}
\end{eqnarray}
The right-handed positions $c_{Ei}$ we fixed by requiring that
with random Yukawa couplings the average charged lepton masses
fit their observed values. The mixings of the left-handed charged
leptons are similar to the neutrino mixings, while the right-handed mixings 
are small.   

The parameter sets (\ref{ps}) and (\ref{psl}) demonstrate that 
bulk fermions in a warped geometry can nicely fit with order
unity parameters not only the huge fermion mass hierarchy but
also the fermion mixings.  Note that in our model a non-trivial wave function  
for fermions is automatically induced by the 
non-factorizable geometry. 

\section{Flavor violation}
There are various sources of flavor violation in the warped model
\cite{HPR,GP,HS2,K00,dAS,KKS02,B02}. 
The low cut-off scale dramatically amplifies the impact of
non-renormalizable operators at the weak scale. With bulk 
fields localized towards the Planck-brane the corresponding 
suppression scales can be significantly enhanced \cite{GP,HS2}.
However, there are limits because the SM fermions need
to have sufficient overlap with the Higgs field at the TeV-brane
to acquire their observed masses. In the case of proton
decay, typical dimension-six operators still have to be suppressed
by small couplings of order $10^{-8}$ to be compatible with
observations \cite{HS2}. 
We consider the following generic four-fermion operators 
which are relevant for flavor violation as well as for proton decay 
\begin{equation}
\int d^4x \int dy \sqrt{-g}\frac{1}{M_5^3}\bar \Psi_i\Psi_j\bar\Psi_k\Psi_l
\equiv \int d^4x \frac{1}{M_4^2}\bar \Psi_i^{(0)}\Psi_j^{(0)}\bar\Psi_k^{(0)}\Psi_l^{(0)}.
\end{equation}
The effective 4D suppression scales $M_4$ associated with these 
operators depend on where the relevant fermion states are localized 
in the extra dimension. Let us focus on some examples. Constraints
on $K-\bar K$ mixing require the dimension-six operator $(ds)^2$
to be suppressed by $M_4>1\cdot 10^{6}$ GeV. Using the fermion positions
of eq.~(\ref{ps}) we find $M_4((ds)^2)=9\cdot 10^7$ GeV, safely above
the experimental bound. The lepton flavor violating decay $\mu\rightarrow eee$
is induced the operator $\mu eee$ at a rate $\Gamma\sim m_{\mu}^5/M_4^4$.
From eq.~(\ref{psl}) we obtain $M_4(\mu eee)=2\cdot 10^7$ GeV, considerably
larger than the experimental bound  $M_4>3\cdot 10^{6}$ GeV. Other
possible dimension-six operators are suppressed in a similar way.

If the SM fermions are located at different positions, KK gauge
bosons couple non-universally to the fermion flavors. The
same is true for the zero modes of Z and W bosons since their
wave functions are $y$-dependent as well \cite{HS,HLS}. The
4D gauge couplings are obtained from an integration over the extra dimension
\begin{equation} \label{9} 
g=\frac{g^{(5)}}{(2\pi R)^{3/2}}\int^{\pi R}_{-\pi R} e^{\sigma}
f_0(y)^2f_A(y)~dy. 
\end{equation} 
Here $f_A$ denotes a generic gauge boson wave function.
Going from the interaction to the mass eigenstates, flavor
non-diagonal couplings are generated as
\begin{equation}
G=U^{\dagger}gU,
\end{equation}
where $g$ is a diagonal matrix in flavor space.
This type of flavor violation
is therefore especially important for large fermion mixing.
The effect is completely analogous to what happens in models 
with family non-universal  Z' bosons, so we can simply adopt
the formalism described, for instance, in ref.~\cite{LP}.  Separating the
fermions in the extra dimension increases the non-universality
of the gauge couplings while suppressing the fermion mixing.
Note that also the right-handed fermion rotations become 
physically relevant. 

In the numerical evaluations we use again
the fermion locations of eqs.~(\ref{ps}) and (\ref{psl}) and average
over random sets of Yukawa couplings. We find a typical value of
Br$(\mu \rightarrow eee) \approx 1\cdot10^{-16}$,
safely below the experimental bound $1\cdot10^{-12}$ \cite{PDG}. The branching ratio
of $\mu\rightarrow e\gamma$ is found to be even six orders
of magnitude below the present experimental sensitivity $1.2\cdot10^{-11}$. 
In the case of muon-electron conversion
in muonic atoms we find Br$(\mu N\rightarrow eN) \approx 1\cdot10^{-16}$
while the current sensitivity is  $6.1\cdot10^{-13}$.
The MECO Collaboration plans to improve this bound by four
orders of magnitude and could therefore reach the
predicted rate. Flavor violation in the quark sector
is also safely suppressed. The $K^0-\bar K^0$ mass splitting, for
instance, induces an upper bound Re$(G_{12}^2)<1\cdot 10^{-8}$ for
the coupling $G_{12}Zds$ \cite{LP} while 
we obtain Re$(G_{12}^2)\approx1\cdot 10^{-12}$. Moreover, we
find Im$(G_{12}^2)\approx3\cdot 10^{-13}$ where CP violation in 
the Kaon system leads to the bound Im$(G_{12}^2)<8\cdot 10^{-11}$. 
Why are the rates for flavor violation so small in the warped model 
whereas in the case of a flat extra dimension they can push the
KK scale up to 5000 TeV   \cite{DPQ}? The reason is that in a
warped geometry the gauge boson wave functions are almost
constant near the Planck-brane \cite{HS,GP,HLS}, where the 
light fermions have to reside in order to explain their small masses.
Therefore the induced deviations from universality are tiny from
the very beginning.

In the scenario of Dirac neutrino masses \cite{GN,HS3} the rate of 
$\mu\rightarrow e\gamma$ transitions is considerably
enhanced by the presence of heavy sterile neutrino states. 
If the SM neutrinos are confined to the TeV-brane,
its large branching ratio pushes the KK scale up to 25 TeV
and thus imposes the most stringent constraint on the model 
\cite{K00}.  
However, the rate for $\mu\rightarrow e\gamma$ is very
sensitive to the mixing between light and heavy neutrino states.
With bulk neutrinos the mixing with heavy states is considerably
reduced. In the case of the large angle MSW solution
we obtain Br$(\mu\rightarrow e\gamma)\approx 10^{-15}$ \cite{HS3}.
While this value is still well below the experimental sensitivity,
it is two orders of magnitude larger than the contribution from gauge
boson exchange and comes close to the reach of the MEG 
experiment \cite{MEG}.

\section{Conclusions} \label{s:concl}
We have shown that bulk quarks and leptons in a warped background 
can naturally accommodate the fermion mass hierarchies and mixings in 
geometrical way, without relying on hierarchical Yukawa couplings. 
Flavor violation by (excited) gauge boson exchange is an immediate 
consequence of this approach, while contributions from non-renormalizable 
operators are automatically suppressed.  Some processes, such as 
muon-electron conversion, are in the reach of next generation experiments
and can provide valuable hints to the higher dimensional theory.

\end{document}